\documentclass{sig-alternate-2013}

\usepackage[hyperfootnotes=false]{hyperref}
\usepackage{graphicx}
\usepackage{cleveref}
\usepackage{verbatim}
\usepackage{amsmath}
\usepackage{xfrac}
\usepackage{subfig}
\usepackage[font=bf,skip=\baselineskip]{caption}
\usepackage{balance}
\usepackage{tabularx}
\usepackage{url}

\usepackage{xspace}

\usepackage[hide]{chato-notes} 

\usepackage{cmm-greek}

\usepackage{longtable}

\usepackage{layouts}
\usepackage{textcomp}
\usepackage{dblfloatfix}
\usepackage{placeins}
\usepackage[numbers,sort]{natbib}
\DeclareRobustCommand{\VAN}[3]{#2}
\usepackage{microtype}

\usepackage{adjustbox}
\usepackage{array}
\usepackage{booktabs}
\usepackage{multirow}
\usepackage{tabularx}
\usepackage{enumitem}

\usepackage{tikz}
\usetikzlibrary{shapes,arrows,calc,positioning}

\usepackage{hyphenat}

\newcommand{\footnoteurl}[1]{\footnote{\url{#1}}}
\newcommand{\para}[1]{\smallskip\noindent\textbf{#1}}

\newcommand{\mytilde}{\raise.30ex\hbox{$\scriptstyle\mathtt{\sim}$}}

\newcommand{\WP}{Wikipedia\xspace}

\newcommand{\ie}{\textit{i.e.}\xspace}
\newcommand{\eg}{\textit{e.g.}\xspace}
\newcommand{\cf}{\textit{cf.}\xspace}
\newcommand{\etal}{\textit{et al.}\xspace}
\newcommand{\vs}{\textit{vs.}\xspace}
\newcommand{\etc}{\textit{etc.}\xspace}

\newcommand{\Secref}[1]{Sec.~\ref{#1}}

\newcommand{\Tabref}[1]{Table~\ref{#1}}

\newcommand{\Figref}[1]{Fig.~\ref{#1}}

\newcommand{\xhdr}[1]{\vspace{1.7mm}\noindent{{\bf #1.}}}

\newcommand\blfootnote[1]{%
  \begingroup
  \renewcommand\thefootnote{}\footnote{#1}%
  \addtocounter{footnote}{-1}%
  \endgroup
}

\begin{document}



\title{Why We Read Wikipedia} 

\makeatletter
\def\@copyrightspace{\@float{copyrightbox}[b]
\begin{center}
\setlength{\unitlength}{1pc}
\begin{picture}(20,4.5) 
\put(0,-0.95){\crnotice{\@toappear}}
\end{picture}
\end{center}
\end@float}
\makeatother


\permission{\copyright 2017 International World Wide Web Conference Committee \\ (IW3C2), published under Creative Commons CC BY 4.0 License.}
\conferenceinfo{WWW 2017,}{April 3--7, 2017, Perth, Australia.}
\copyrightetc{ACM \the\acmcopyr}
\crdata{978-1-4503-4913-0/17/04. \\
	http://dx.doi.org/10.1145/3038912.3052716 \\
	\includegraphics[scale=0.8]{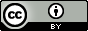}}

\author{
\alignauthor
Philipp Singer\textsuperscript{\textasteriskcentered 1},
Florian Lemmerich\textsuperscript{\textasteriskcentered 1},
Robert West\textsuperscript{\textdagger 2},\\
Leila Zia\textsuperscript{3},
Ellery Wulczyn\textsuperscript{3},
Markus Strohmaier\textsuperscript{1},
Jure Leskovec\textsuperscript{4}\\
\vspace{2mm}
\affaddr{\textsuperscript{1}GESIS \& University of Koblenz-Landau},
\affaddr{\textsuperscript{2}EPFL},
\affaddr{\textsuperscript{3}Wikimedia Foundation},
\affaddr{\textsuperscript{4}Stanford University}\\
\textsuperscript{1}firstname.lastname@gesis.org,
\textsuperscript{2}robert.west@epfl.ch,
\textsuperscript{3}firstname@wikimedia.org,
\textsuperscript{4}jure@cs.stanford.edu\\
}

\maketitle

\blfootnote{\textsuperscript{\textasteriskcentered}Both authors contributed equally to this work.}
\blfootnote{\textsuperscript{\textdagger}Robert West is a Wikimedia Foundation Research Fellow.}
\vspace{-7mm}
\begin{abstract}
Wikipedia is one of the most popular sites on the Web, with millions of users relying on it to satisfy a broad range of information needs every day.
Although it is crucial to understand what exactly these needs are in order to be able to meet them, little is currently known about why users visit Wikipedia.
The goal of this paper is to fill this gap by combining a survey of Wikipedia readers with a log-based analysis of user activity.
Based on an initial series of user surveys, we build a taxonomy of Wikipedia use cases along several dimensions, capturing users' motivations to visit Wikipedia, the depth of knowledge they are seeking, and their knowledge of the topic of interest prior to visiting Wikipedia.
Then, we quantify the prevalence of these use cases via a large-scale user survey conducted on live Wikipedia with almost 30,000 responses. Our analyses highlight the variety of factors driving users to Wikipedia, such as current events, media coverage of a topic, personal curiosity, work or school assignments, or boredom.
Finally, we match survey responses to the respondents' digital traces in Wikipedia's server logs, enabling the discovery of behavioral patterns associated with specific use cases.
For instance, we observe long and fast-paced page sequences across topics for users who are bored or exploring randomly, whereas those using Wikipedia for work or school spend more time on individual articles focused on topics such as science.
Our findings advance our understanding of reader motivations and behavior on Wikipedia and can have implications for developers aiming to improve Wikipedia's user experience, editors striving to cater to their readers' needs, third-party services (such as search engines) providing access to Wikipedia content, and researchers aiming to build tools such as recommendation engines.

\end{abstract}


\vspace{1mm}

\section{Introduction}
\label{sec:intro}

Wikipedia is the world's largest encyclopedia and one of the most popular websites, with more than 500 million pageviews per day. It attracts millions of readers from across the globe and serves a broad range of their daily information needs.
Despite this, very little is known about the motivations and needs of this diverse user group: why they come to Wikipedia, how they consume the content in the encyclopedia, and how they learn. Without this knowledge, creating more content, products, and services that ensure high levels of user experience remains an open challenge \cite{basu2003context,white2009characterizing,feild2010predicting,krug2014make}.

\para{Background and objectives.}
A rich body of work has investigated motivations and behavior patterns of users on the Web \cite{goel2012does,kumar2010characterization}. Specific attention has been cast on a few major sites including search engines \cite{broder2002taxonomy,rose2004understanding,weber2011uses}, and social networking sites such as Twitter \cite{java2007we,kwak2010twitter} and Facebook \cite{ryan2011uses}. Yet, surprisingly little is known about the motivations, needs, and behaviors of \WP readers, possibly keeping \WP from reaching its full potential. 

The vast literature on user behavior in \WP (\cf\ Okoli \etal\ \cite{okoli2012people} for an overview) has focused on content production. It mainly investigates editors' motivations \cite{arazy,nov2007motivates}, patterns of editing behavior \cite{jurgens2012temporal}, and the quality of content \cite{kittur2008harnessing,stvilia2008information}. Much less is known about content consumption, even though readers make up the majority of Wikipedia users. The limited work on readers has focused on topics such as content preferences \cite{spoerri2007popular,ratkiewicz2010characterizing,lehmann2014reader}, search queries leading to \WP \cite{waller2011search}, or navigation patterns \cite{paranjape2016improving,lamprecht2016evaluating,west2012human,singer2014detecting}. In contrast, the present work aims at understanding \emph{why we read Wikipedia}.

\begin{figure}[t!]
	\centering
	\includegraphics[width=0.3\textwidth]{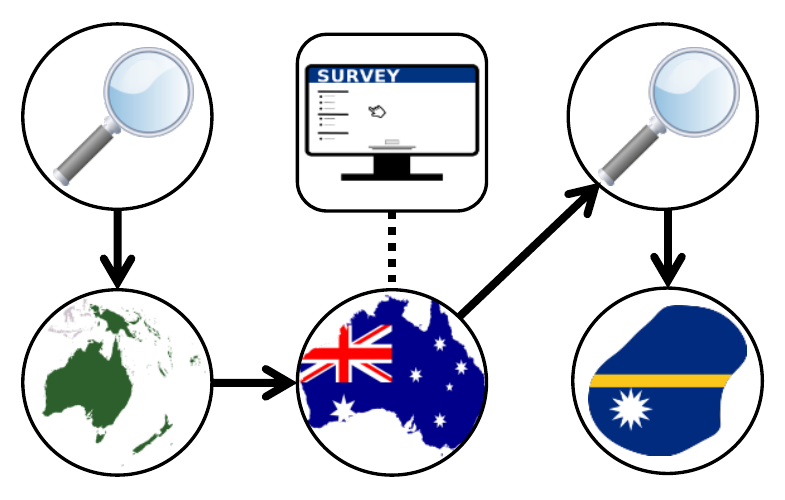}
	\caption{
	\emph{Example Wikipedia reading session.}
	The user arrives from a search engine and visits the article about \textit{Oceania;} she then navigates to \textit{Australia,} where she responds to our survey. Afterwards, the reader goes back to the search engine and finally visits the article about \textit{Nauru}. This paper studies survey responses as well as webrequest logs.}
	\label{fig:fig1}
\end{figure}

\para{Materials and methods.}
We present a robust taxonomy of use cases for reading Wikipedia, constructed through a series of surveys based on techniques from grounded theory \cite{strauss1998basics}. Initially, we administered a survey to elicit free text responses to the question, \emph{Why are you reading this article today?} Based on the responses, we designed a taxonomy covering three major dimensions that can be used to characterize the observed use cases.
After validating the robustness of our taxonomy, we study the prevalence of use cases as measured by a large-scale multiple\hyp choice survey on English Wikipedia. To correct for various forms of representation bias in our pool of respondents, we use inverse propensity score weighting adjustment. We then enrich the survey data by linking each survey response to the respondent's behavior traces mined from \WP's webrequest logs. An illustration of how both survey and log data are collected can be found in \Figref{fig:fig1}. Finally, we use the joined survey and log data to identify characteristic behavior patterns for reader groups with specific intentions via subgroup discovery \cite{kloessgen1996,herrera2010}.

\para{Contributions and findings.}
The following are our three main contributions:
(i)~We present a robust taxonomy for character\-izing use cases for reading Wikipedia (\Secref{sec:taxonomy}), which captures users' motivations to visit Wikipedia, the depth of information they are seeking, and their familiarity with the topic of interest prior to visiting Wikipedia.
(ii)~We quantify the prevalence and interactions between users' motivations, information needs, and prior familiarity via a large-scale survey yielding almost 30,000 responses (\Secref{sec:survey_results}). 
(iii)~We enhance our understanding of the behavioral patterns associated with different use cases by combining survey responses with digital traces recorded in Wikipedia's server logs  (\Secref{sec:webrequest_results}). 

Our analysis lets us conclude that there is a variety of motivations bringing readers to Wikipedia, which can be characterized by distinct behavioral patterns. For example, users visiting \WP out of boredom view multiple, topically different articles in quick succession.
While no motivation clearly dominates, it is generally the case that Wikipedia is used for shallow information needs (fact look-up and overview) more often than for in-depth information needs. 
Also, prior to reading an 
article, readers are familiar with its topic about as often as not.

The outcomes of this research can help Wikipedia's editor and developer communities, as well as the Wikimedia Foundation, to make more informed decisions about how to create and serve encyclopedic content in the future.

\section{\hspace{-0.3mm}Taxonomy of Wikipedia readers}

\label{sec:taxonomy}
Our research relies on a taxonomy of Wikipedia readers, something that was previously absent from the literature. We designed and analyzed a series of surveys based on techniques from \emph{grounded theory} \cite{strauss1998basics} to build a robust categorization scheme for Wikipedia readers' motivations and needs.
In this section, we explain the individual steps taken and the resulting taxonomy.


\para{Building the initial taxonomy.}
We started with an initial questionnaire \emph{(Survey 1)}, where a randomly selected 
subgroup of English Wikipedia readers (sampling rate 1:200, desktop and mobile, 4 days, about 5,000 responses) saw a survey widget while browsing Wikipedia articles. If the reader chose to participate, she was taken to an external site (Google Forms) and asked to answer the question \textit{``Why are you reading this article today?"}\ in free-form text (100\hyp character limit). 

To arrive at categories for describing use cases of Wikipedia reading, five researchers performed three rounds of hand\hyp coding on a subset of the 5,000 responses, without discussing any expectations or definitions ahead of time.
In the first stage, all researchers worked together on 20 entries to build a common understanding of the types of response.
In the second stage, based on the discussions of the first stage, tags were generously assigned by each researcher individually to 100 randomly selected responses, for a total of 500 responses tagged. All 500 tagged responses were reviewed, and four main trends (motivation, information need, context, and source) were identified, alongside tags associated with each response.
In the third stage, each researcher was randomly assigned another 100 responses and assessed if they contained information about the four main trends identified in the previous stage and if the trends and tags should be reconsidered. 
The outcome of these stages revealed the following three broad ways in which users interpreted the question; we use them as orthogonal dimensions to shape our taxonomy: 

\begin{itemize}[leftmargin=*, topsep=1pt,itemsep=-1ex,partopsep=1ex,parsep=1ex]
\item \emph{Motivation}: work/school project, personal decision, current event, media, conversation, bored/random, intrinsic learning.
\item \emph{Information need}: quick fact look-up, overview, in-depth.
\item \emph{Prior knowledge}: familiar, unfamiliar.
\end{itemize}

\para{Assessing the robustness of the taxonomy.}
We conducted two surveys similar to \emph{Survey 1} on the Spanish and Persian Wikipedias which resulted in similar observations and dimensions as above.
Additionally, we assessed the robustness of the above taxonomy in two follow-up surveys.
First, we ran a survey identical to Survey~1 to validate our categories on unseen data
(\emph{Survey 2;} sampling rate 1:2000, mobile, 3 days, 1,650 responses).
No new categories were revealed through hand\hyp coding.


Second, we crafted a multiple\hyp choice version of the free\hyp form surveys (\emph{Survey 3;} sampling rate 1:200, desktop and mobile, 6 days, about 10,500 responses).
It comprised three questions with the following answer options in random order (the first two questions also offered ``other'' as an answer, with the option to enter free\hyp form text):

\begin{itemize}[leftmargin=*, topsep=1pt,after=\vspace{1pt},itemsep=-1ex,partopsep=1ex,parsep=1ex]
\item \emph{I am reading this article because\dots}: I have a work or school-related assignment; I need to make a personal decision based on this topic (\eg, buy a  book, choose a travel destination); I want to know more about a current event (\eg, a soccer game, a recent earthquake, somebody's death); the topic was referenced in a piece of media (\eg, TV, radio, article, film, book); the topic came up in a conversation; I am bored or randomly exploring Wikipedia for fun; this topic is important to me and I want to learn more about it (\eg, to learn about a culture). 
Users could select multiple answers for this question.
\item \emph{I am reading this article to\dots}: look up a specific fact or to get a quick answer; get an overview of the topic; get an in-depth understanding of the topic. 
\item \emph{Prior to visiting this article\dots}: I was already familiar with the topic; I was not familiar with the topic, and I am learning about it for the first time. 
\end{itemize}

Only 2.3\% of respondents used the ``other'' option, and hand-coding of the corresponding free\hyp form responses did not result in new categories.
We thus conclude that our categories are robust and use the resulting classification as our \emph{taxonomy of Wikipedia readers} in the rest of this paper.


\begin{table*}[t!]
\centering
\small
\caption{\emph{Features.} This table describes all features utilized in this work. \emph{Survey features} capture responses to our survey questions; \emph{request features} capture background information about the respondent mined from webrequest logs; \emph{article features} describe the requested Wikipedia article; and \emph{session\slash activity features} are derived from the entire reading session and beyond\hyp session activity.
}
\vspace{-1em}
\begin{tabularx}{0.99\textwidth}{l|l|X} 
\toprule 
 & feature & description \\ 
\midrule
& motivation & Type of motivation for reading an article, as selected by respondent in survey. As multiple responses were allowed, we work with boolean dummy variables for each motivation.\\
\raisebox{.8\normalbaselineskip}[0pt][0pt]{\rotatebox[origin=c]{90}{survey~}} & information need & Information need for reading an article, as selected by respondent in survey.\\ 
& prior knowledge & Prior knowledge about the topic before visiting the article, as selected by respondent in survey.\\ \hline
& country & Country code of respondent in survey (e.g., USA) derived from the IP address. \\
  & continent & Continent of respondent in survey (e.g., North America) derived from the IP address. \\ 
& local time weekday & Local weekday of survey request detected by timezone information (Mon-Sun).  \\
\raisebox{1\normalbaselineskip}[0pt][0pt]{\rotatebox[origin=c]{90}{request~~~}} & local time hour & Local hour of survey request detected by timezone information (0-24).\\
 & host & Requested Wikipedia host: ``desktop'' (en.wikipedia.org), or ``mobile web'' (en.m.wikipedia.org). \\
& referer class & Referer class of request (none, internal, external, external search engine, or unknown).\\ 
\hline
& article in-degree & The topological in-degree of an article. \\
& article out-degree & The topological out-degree of an article. \\
& article pagerank & The unnormalized pagerank of an article; calculated with damping factor of 0.85. \\
& article text length & The text length of an article as extracted from HTML---number of characters. \\
& article pageviews & The sum of pageviews for the article in same time period as survey. \\
\raisebox{0\normalbaselineskip}[0pt][0pt]{\rotatebox[origin=c]{90}{article~~~}} & article topics & Probability vector for 20 topics as extracted by LDA from a bag-of-words representation. Topics were manually labeled as follows: (1) transportation \& modern military, (2) biology \& chemistry, (3) South Asia, Middle East, (4) mathematics, (5) 21st century, (6) TV, movies, \& novels, (7) Britain \& Commonwealth, (8) East Asia, (9) Spanish (stubs), (10) war, history, (11) geography (unions, trade), (12) literature, art, (13) education, government, law, (14) 20th century, (15) sports, (16) United States, (17) numbers, (18) technology, energy, \& power, (19) music, and (20) geographical entities. We use the probabilistic topic distribution as 20 individual features for each article. \\
& article topic entropy & Measures the topic specificity of an article from LDA probability vector. \\ \hline
& session length & The number of requests within the session. \\
& session duration & Total time spent in the session in minutes. \\
 & avg. time difference & Average time difference between subsequent session requests (i.e., dwelling time). \\
& avg. pagerank difference & Average pagerank difference between subsequent session requests (i.e., stating whether readers move to periphery or core).\\
 & avg. topic distance & Average Manhattan distance between topic distributions for subsequent session requests (i.e., capturing topical changes). \\
\raisebox{-0\normalbaselineskip}[0pt][0pt]{\rotatebox[origin=c]{90}{~~~~~session/activity}} & referer class frequency & For each referer class (see survey features): frequency in session. \\
& session article frequency & The number of times requested article for survey response occurs within the session. \\
& session position & Relative position inside a session when answering the survey. \\
& num. of sessions & The total number of sessions for respondent in survey time period. \\
& num. of requests & The total number of requests for respondent in survey time period. \\
\bottomrule 
\end{tabularx} 
\label{tab:features}
\vspace{-1em}
\end{table*}

\newpage
\section{Datasets and Preprocessing}
\label{sec:data+methodology}

Here, we describe utilized datasets and preprocessing.


\subsection{Survey}
\label{sec:survey_description}
To quantify the prevalence of the driving factors specified by our taxonomy, we ran an additional \emph{large-scale survey} on English Wikipedia consisting of the same three questions on motivation, depth of information need, and prior knowledge as \emph{Survey 3} (\Secref{sec:taxonomy}).
The survey was run at a sampling rate of 1:50
from 2016-03-01 to 2016-03-08 on all requests to English Wikipedia's mobile and desktop sites.
It was not shown on non\hyp article pages (discussion pages, search pages, \etc), on the main page of Wikipedia, and to browsers with \emph{Do not Track} enabled. Potential survey participants were identified by assigning a token to their browsers and eventually showing a widget with an invitation to participate in the survey. 
Once shown, the reader could ignore it, dismiss it, or opt in to participate which would take the reader to an external site (Google Forms), where she would see the three questions described in \Secref{sec:taxonomy}. A unique, anonymous ID was passed to Google Forms for each user, which would later be used to link the survey responses to users' webrequest logs (\Secref{sec:webrequest_logs}).
A privacy and consent statement\footnote{\url{https://wikimediafoundation.org/wiki/Survey_Privacy_Statement_for_Schema_Revision_15266417}} providing details about the collection, sharing, and usage of the survey data was shown to all users prior to submitting their responses.
Overall, our dataset consists of survey answers from 29,372 participants after basic data cleaning such as removing duplicate answers from the same users.
Whenever we write ``survey'' throughout the rest of this paper, we refer to the survey described here.

\subsection{Webrequest logs}
\label{sec:webrequest_logs}

Ultimately, we aim to to understand how users' motivation, desired depth of knowledge, and prior knowledge (\ie, their answers to our survey) manifest themselves in their reading behavior.
The data collected through the survey alone, however, does not provide any information on the respondent's behavior beyond the single pageview upon which the survey was presented.

In order to be able to analyze respondents' reading behavior in context, we connect survey responses to the webrequest logs maintained by \WP's web servers, where every access to any \WP page is stored as a record that contains, among others, the requested URL, referer URL, timestamp, client IP address, browser version, and city\hyp level geolocation inferred from the IP address.
Since the logs do not contain unique user IDs, we construct approximate user IDs by concatenating the client IP address and browser version; \cf\ discussion in \Secref{sec:Reflections on methodology}.

As the information needs and reading behavior of the same user may change over time, we operate at an intermediate temporal granularity by decomposing a user's entire browsing history into \emph{sessions,} where we define a session as a contiguous sequence of pageviews with no break longer than one hour \cite{halfaker2015user}.
To reconstruct the session in which a user took our survey (\cf\ \Figref{fig:fig1}), we retrieved from the webrequest logs all records with the user's (approximate) ID, ordered them by time, chunked them into sessions according to the aforementioned one-hour rule, and returned the session that contains the record with the specific URL and timestamp of the survey response.

\begin{figure*}[t!]
	\centering
	\subfloat{\includegraphics[width=0.99\textwidth]{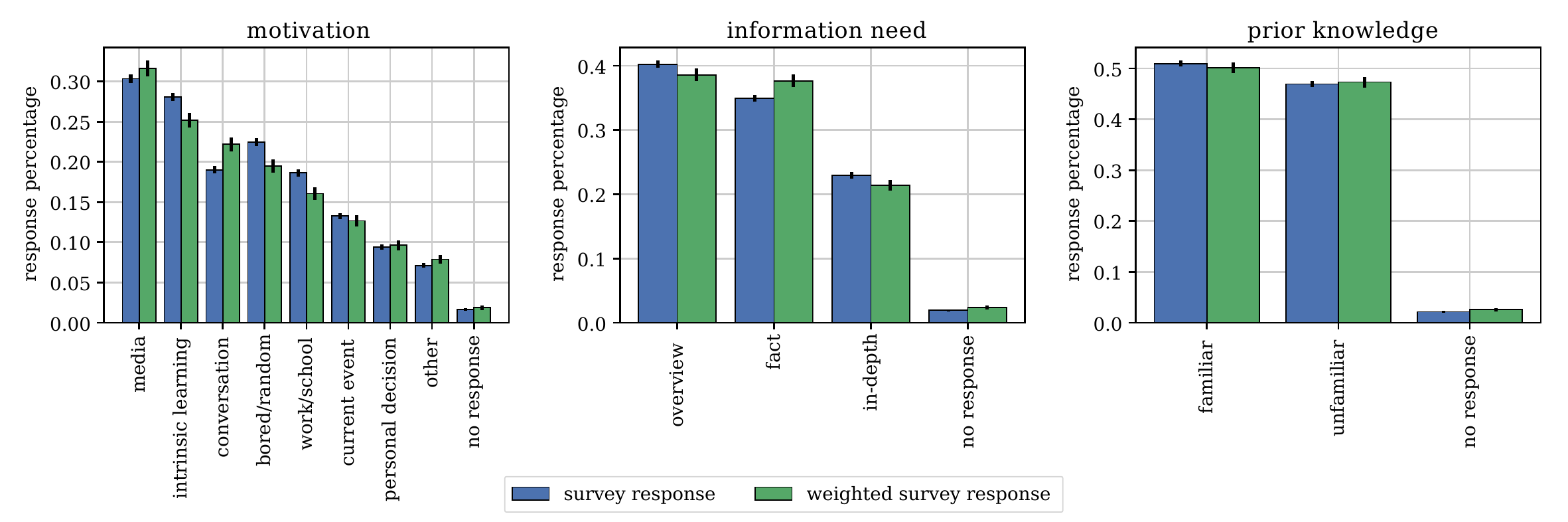}\label{<figure1>}}
		\vspace{-1em}
	\caption{\emph{Survey responses.} This figure visualizes the share of answers for the three parts of the user survey: motivation, information need, and prior knowledge. The blue bars (left) reflect the raw unweighted responses, and the green bars (right) depict the bias-corrected weighted responses
	(propensity score weight adjustment).
	Error bars visualize the 95\% confidence intervals using effective sample size for the weighted responses. In general, results suggest popularity of both extrinsic and intrinsic motivation, as well as high, but balanced,  relevance for certain information needs and prior knowledge.
	The results based on bias-correction weighting reflect minor changes in responses without drastically changing the general direction. }
	\label{fig:survey_response}
	\vspace{-1em}
\end{figure*}

\subsection{Wikipedia article data}
\label{sec:wikipedia_background_data}


Different articles are consumed in different ways.
Hence, the properties of articles viewed by survey respondents play an important role in our analysis.
To extract these properties, we utilized the public dump of English \WP released on 2016-03-05,%
\footnoteurl{https://archive.org/details/enwiki-20160305}
such that article revisions closely match those seen by survey participants.

The dump contains wiki markup, whereas browsers receive HTML code generated from this markup. Since the markup may contain templates that are expanded only upon conversion to HTML, some page content is not immediately available in markup form.
In order to obtain a more complete representation, we retrieved the full HTML of the article contents using Wikimedia's public API. %
 In addition to the textual article content, we extracted the network of articles (5.1M) connected by hyperlinks (370M).



\subsection{Features}
\label{sec:features}

Throughout this work, we study features extracted 
directly from the survey responses (\Secref{sec:survey_description}), from underlying webrequest logs of article requests and extracted sessions (\Secref{sec:webrequest_logs}), and background Wikipedia article data associated with requested articles (\Secref{sec:wikipedia_background_data}). We list and describe all features utilized in this work in \Tabref{tab:features}.

For topic detection, we fit a Latent Dirichlet Allocation (LDA) \cite{blei2003latent} model on bag-of-words vectors  representing articles' textual content (with stopwords removed) using online variational Bayes.
To find a balance between complexity and interpretability, we decided to work with 20 topics. We assigned labels to topics by manually inspecting the topics' word distributions and their top Wikipedia articles.

\subsection{\hspace{-1mm}Correcting survey bias via webrequest logs}
\label{sec:correction}

The goal of this work is to study the motivations and behaviors representative of Wikipedia's entire reader population.
However, deducing properties of a general population from surveying a limited subpopulation is subject to different kinds of biases and confounders, including
\emph{coverage bias} (inability to reach certain subpopulations), \emph{sampling bias} (distortions due to sampling procedure),
and \emph{non\hyp response bias} (diverse likelihood of survey participation after being sampled as a participant).
%

Consequently, an important step in our analysis is to account for potential biases in survey responses.
Finding suitable adjustments has been a decade\hyp long research effort in the survey methodology community \cite{brick2013unit}. Since strata methods such as poststratification~\cite{gelman2000poststratification} are less well suited to control for a large number of features, we opt for \emph{inverse propensity score weighting} \cite{austin2011introduction} as an alternative.
This technique assigns control weights to each survey response, thus correcting bias with respect to a control group (Wikipedia population). The rationale behind this procedure is that answers of users less likely to participate in the survey should receive higher weights, as they represent a larger part of the overall population with similar features. For determining participation probabilities \emph{(propensity scores),} we use gradient\hyp boosted regression trees on individual samples to predict if they belong to the survey \vs\ the control group, using all features of \Secref{sec:features}.
(We provide additional methodological details in the appendix.)
By using background features (\eg, country, time) plus digital traces (\eg, sessions),
and by building a representative control group,
we have an advantage over traditional survey design, which is often limited to few response features such as gender and age, as well as to small control groups.

When discussing results in the next section, we shall see (\Figref{fig:survey_response}) that our weight adjustment changes the relative shares of survey responses only slightly, with general trends staying intact.
Hence, we shall utilize only weighted survey responses for inference on statistical properties from this point on. Additionally, we use the so-called \emph{effective sample size} (\cf\ appendix) when calculating standard errors, confidence intervals, and statistical tests, in order to account for differing standard errors of weighted estimators.

\section{Results: Why we read Wikipedia}

This section discusses results on why users read Wikipedia.


\subsection{Survey results}
\label{sec:survey_results}

We start with a discussion of the responses to our survey. 

\para{Survey responses.}
First, we examine the percentages of survey respondents with specific motivations, information needs, and prior knowledge.
We visualize the results in \Figref{fig:survey_response}, focusing on the green (right) bars representing weighted survey responses (sorted by popularity).

With respect to \emph{motivation,} we find that Wikipedia is consulted in a large spectrum of use cases and that no clearly dominant motivation can be identified. Prominently, extrinsic situations trigger readers to visit Wikipedia to look up a topic that was referenced in the media (30\%), came up in a conversation (22\%), is work or school\hyp related (16\%), or corresponds to a current event (13\%). At the same time, readers have intrinsic motivations, such as wanting to learn something (25\%), being bored (20\%), or facing a personal decision (10\%).
We also find that the ``other'' option was only rarely selected, further confirming the robustness of the taxonomy of readers introduced in \Secref{sec:taxonomy}.

The results also show that Wikipedia is visited to satisfy different kinds of \emph{information needs}. 
Interestingly, shallow information needs (overview [39\%] and quick fact-checking [38\%]) appear to be more common than deep information needs (21\%).
As for \emph{prior knowledge,} we observe nearly identical shares of readers being familiar (50\%) \vs\ unfamiliar (47\%) with the topic of interest.




\begin{table}[b!]
\vspace{-1.5em}
\small
	\centering
	\caption{\emph{Survey response correlations.} 
		Each cell depicts the row-normalized share of responses that have also selected a given column as answer (without ``other'' and non\hyp responses). The bottom rows highlight the overall share of responses for a given column as expectation. Values in brackets reflect the lift ratio of observed \vs\ expected frequency.
		The last column indicates significance (***~$<$~0.001, **~$<$~0.01, *~$<$~0.05) for the hypothesis test of independence of observed frequencies (contingency table with row frequencies and complement of all other rows) and expected frequencies (as in the last table row) using a $\chi^2$ test using the effective sample size.}
	\subfloat[Motivation vs. information need]{\begin{tabularx}{0.49\textwidth}{l|XXX|l} 
\toprule 
information need & fact & in-depth & overview & sig. \\ 
motivation &  &  &  &  \\ 
\midrule 
media & 0.38 (1.00) & 0.19 (0.87) & 0.43 (1.12) & *** \\ 
intrinsic learning & 0.29 (0.76) & 0.35 (1.62) & 0.35 (0.92) & *** \\ 
conversation & 0.43 (1.13) & 0.20 (0.93) & 0.36 (0.94) & *** \\ 
bored/random & 0.31 (0.83) & 0.23 (1.05) & 0.45 (1.17) & *** \\ 
work/school & 0.39 (1.04) & 0.23 (1.09) & 0.36 (0.93) &  \\ 
current event & 0.36 (0.95) & 0.28 (1.30) & 0.35 (0.92) & *** \\ 
personal decision & 0.32 (0.85) & 0.29 (1.35) & 0.38 (0.97) & *** \\ 
\hline 
response perc. & 0.38 & 0.21 & 0.39 & \\ 
\bottomrule 
\end{tabularx} 
\label{table:cross1}} \\
	\subfloat[Motivation vs. prior knowledge]{\begin{tabularx}{0.45\textwidth}{l|XX|l} 
\toprule 
prior knowledge & familiar & unfamiliar & sig. \\ 
motivation &  &  &  \\ 
\midrule 
media & 0.42 (0.83) & 0.58 (1.22) & *** \\ 
intrinsic learning & 0.57 (1.14) & 0.41 (0.87) & *** \\ 
conversation & 0.49 (0.98) & 0.49 (1.04) & *** \\ 
bored/random & 0.53 (1.07) & 0.45 (0.95) &  \\ 
work/school & 0.52 (1.04) & 0.46 (0.97) &  \\ 
current event & 0.52 (1.03) & 0.46 (0.98) &  \\ 
personal decision & 0.50 (0.99) & 0.48 (1.02) &  \\ 
\hline 
response perc. & 0.50 & 0.47 & \\ 
\bottomrule 
\end{tabularx} 
\label{table:cross2}} \\
	\subfloat[Prior knowledge vs. 
	information need ]{\begin{tabularx}{0.45\textwidth}{l|XXX|l} 
\toprule 
information need & fact & in-depth & overview & sig. \\ 
prior knowledge &  &  &  &  \\ 
\midrule 
familiar & 0.43 (1.13) & 0.25 (1.15) & 0.32 (0.83) & *** \\ 
unfamiliar & 0.34 (0.90) & 0.19 (0.87) & 0.47 (1.22) & *** \\ 
\hline 
response perc. & 0.38 & 0.21 & 0.39 & \\ 
\bottomrule 
\end{tabularx} 
\label{table:cross3} }
	\label{table:cross}
	\vspace{-1em}
\end{table}

\para{Survey response correlations.}
Next, in \Tabref{table:cross}, we study whether certain combinations of motivations, information needs, and prior knowledge occur more frequently than expected, quantified by the \emph{lift,} \ie, the ratio between observed and expected frequencies.

\Tabref{table:cross1} suggests that 
different \emph{motivations} are coupled with different \emph{information depths}.
Specifically, in-depth information needs prevail when readers are driven by intrinsic learning (lift $1.62$);
quick fact look-ups are associated more strongly with conversation and work\slash school motivations than one would expect \textit{a priori;}
and gaining an overview of a topic appears to be especially important for readers motivated by media coverage and for the bored.



In \Tabref{table:cross2}, we find weaker correlations between \emph{motivations} and levels of \emph{prior knowledge,} apparent from lifts closer to 1 and a lack of significance. However, certain trends still emerge; \eg, when readers research a topic from the media, they are more likely to be unfamiliar with the topic (lift $1.22$). In contrast, readers whose goal is learning are more likely to be familiar with the topic (lift $1.14$).

As a corollary of the above correlations, we also observe patterns when contrasting \emph{prior knowledge} with \emph{information need} (\Tabref{table:cross3}). We find that familiar readers are more likely to look up quick facts (lift $1.13$) and aim at getting in-depth knowledge about a topic (lift $1.15$) than one would expect. Contrarily, unfamiliar readers are more likely to first aim at getting an overview of the topic (lift $1.22$) instead of directly going into depth (lift $0.87$).

\begin{figure}[t!]
	\centering
	\subfloat{\includegraphics[width=0.49\textwidth]{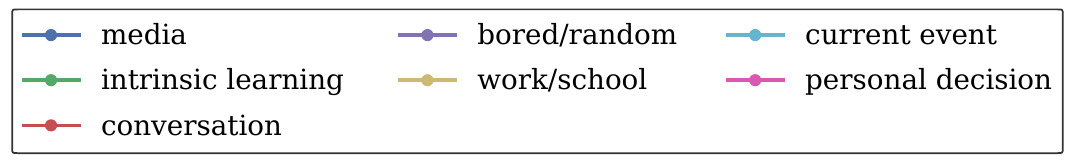}} \\ \vspace{-1.4em}
	\addtocounter{subfigure}{-1}
\subfloat{\includegraphics[width=0.49\textwidth]{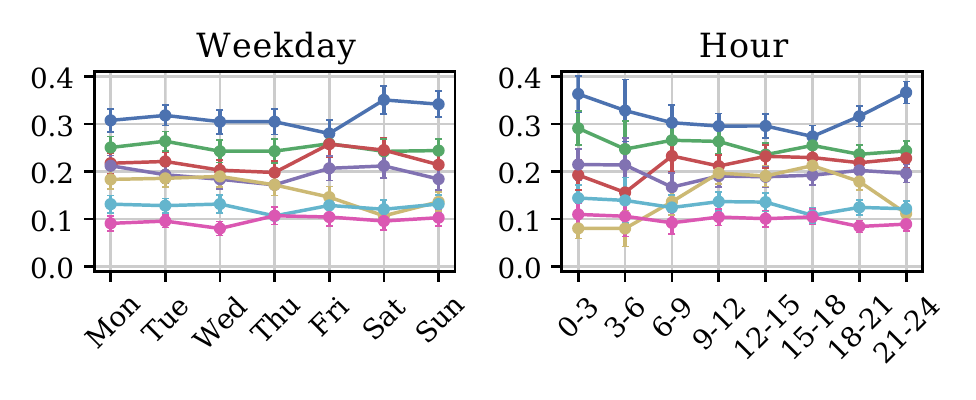}\label{fig:weekday_hour}}
	\vspace{-1em}
	\caption{\emph{Motivation day and time.} This figure visualizes how the relative share of motivation (y-axes) changes over the course of a week and the course of a day (x-axes). 
	Error bars depict the 95\% confidence interval with effective sample size. 
	}
	\vspace{-2em}
	\label{fig:time}
\end{figure}

\para{Survey responses over time.}
Next, we study how the prevalence of motivations, information needs, and prior knowledge changes over time. 
For \emph{motivations,} shown in \Figref{fig:time}, we find relatively stable trends over the course of a week or day.
Three notable exceptions, however, emerge.
First, on weekends (Saturday, Sunday) and at night, there is a higher share of readers who are led to Wikipedia by media coverage; this is potentially due to a higher likelihood of being exposed to media during these time periods.
Similarly, conversations play a more important role on Fridays and Saturdays, possibly since people go out, meet with friends, and are involved in conversations that lead to consulting Wikipedia.
By contrast, reading an article for work or school reasons has a relatively lower share towards the weekend, but peaks at daytime hours, probably because people work and go to school on working days and during daytime hours.

Additionally, results on \emph{information need} show overall quite stable trends over a week and over a day
without clear outliers,
also due to larger confidence intervals (results not visualized). For \emph{prior knowledge,} we identify small upward trends on weekends and evening hours for already being familiar with the topic, compared to being unfamiliar. However, error bars are again too large to justify stronger claims.

%

\begin{table*}[t!]
	\small
	\centering
\caption{\emph{Top subgroups for the motivations ``work/school'' and ``bored/random''.} Each  table shows the top subgroups with significantly different shares of users with a certain motivation $T$. 
For each subgroup $S$, we display
the relative size $P(S)$ of the subgroup (\ie, the share of users covered by the subgroup description),
the share $P(S|T)$ of the subgroup among those with motivation $T$,
the target share $P(T|S)$ in the subgroup,
and the \emph{lift} measure, defined as $P(T|S)/P(T) = P(S|T)/P(S)$.
Rows are ranked by lift.
The last column indicates significance (***~$<$~0.001, **~$<$~0.01, *~$<$~0.05) for the hypothesis test of independence between subgroup and target motivation using a $\chi^2$ test with effective sample size and Bonferroni correction.}
\vspace{-1em}
	\fontsize{6.5}{\baselineskip}\selectfont
\subfloat[$T$: ``motivation = work/school''; $P(T) = 16.1\%$]{
\begin{tabularx}{0.5\textwidth}{l|XXXX|l} 
\toprule
                                Subgroup $S$ &  $P(S)$ & $P(S|T)$ & $P(T|S)$ &  lift & sig. \\
\midrule
topic (mathematics) &  7.9\% &  17.1\% &  34.8\% &  2.17 &  *** \\
topic (war, history) &  4.4\% &   9.6\% &  34.7\% &  2.16 &  *** \\
topic (technology) & 13.2\% &  23.7\% &  28.8\% &  1.79 &  *** \\
topic (biology, chemistry) &  8.6\% &  14.0\% &  26.2\% &  1.63 &  *** \\
host = desktop & 35.5\% &  57.8\% &  26.1\% &  1.63 &  *** \\
article pagerank $\geq$ 9.98 & 20.0\% &  32.4\% &  26.1\% &  1.62 &  *** \\
avg. time difference $\geq$ 9.40 &  7.7\% &  11.5\% &  24.0\% &  1.50 &  *** \\
avg. pagerank difference < -4.35 &  7.6\% &  11.2\% &  23.6\% &  1.47 &  *** \\
topic (literature, art) & 10.1\% &  14.7\% &  23.5\% &  1.46 &  *** \\
avg. time difference: [3.60:9.40[ &  7.7\% &  11.0\% &  23.1\% &  1.44 &  *** \\
num. (referer=search) $\geq$ 2 & 20.5\% &  28.5\% &  22.4\% &  1.39 &  *** \\
session duration $\geq$ 6.60 & 18.0\% &  24.2\% &  21.6\% &  1.34 &  *** \\
\bottomrule
\end{tabularx}
}
\subfloat[$T$: ``motivation = bored/random''; $P(T) = 19.5\%$] {
\begin{tabularx}{0.5\textwidth}{l|XXXX|l} 
\toprule
Subgroup $S$ &  $P(S)$ & $P(S|T)$ & $P(T|S)$ &  lift & sig. \\
\midrule
referer class: internal &  9.4\% &  14.0\% &  29.0\% &  1.49 &  *** \\
num. of requests $\geq$ 8 & 11.8\% &  16.6\% &  27.5\% &  1.41 &  *** \\
topic (sports) &  5.9\% &   8.0\% &  26.1\% &  1.34 &   ** \\
num. (referer=internal) $\geq$ 1 & 17.1\% &  22.7\% &  25.9\% &  1.33 &  *** \\
session position: [0.33:0.75[ &  7.5\% &   9.8\% &  25.6\% &  1.31 &   ** \\
avg. topic distance  $\geq$ 1.08 &  7.5\% &   9.8\% &  25.2\% &  1.29 &    * \\
topic (21st century) & 25.1\% &  32.1\% &  25.0\% &  1.28 &  *** \\
session length $\geq$ 3 & 22.2\% &  28.3\% &  24.8\% &  1.27 &  *** \\
avg. time difference: [0.68:1.56[ &  7.7\% &   9.7\% &  24.7\% &  1.27 &    * \\
num. (referer=none) $\geq$ 2 &  9.7\% &  12.2\% &  24.5\% &  1.26 &    * \\
topic (tv, movies, novels) & 34.1\% &  41.4\% &  23.7\% &  1.21 &  *** \\
\# article pageviews $\geq$ 63606 & 19.8\% &  23.5\% &  23.1\% &  1.19 &   ** \\
\bottomrule
\end{tabularx}
}
\label{tab:sgResults}
\vspace{-2em}
\end{table*}

\subsection{Webrequest\hyp log results}
\label{sec:webrequest_results}
Our previous results suggest that Wikipedia is visited for a variety of use cases that differ not only in their motivation triggers, but also in the depth of information needs, and readers' prior familiarity with the topic.
In this section, we investigate correlations of survey responses with behavioral patterns based on request, article, and session features (\Secref{sec:features}).
In doing so, we reveal characteristic differences and develop stereotypes for motivational groups.




\para{Methodology. }
Due to our large set of features at interest (\Secref{sec:features}), we investigate behavioral reader patterns based on 
 rule mining techniques, specifically \emph{subgroup discovery}~\cite{kloessgen1996,herrera2010}.
The general goal of subgroup discovery is to find descriptions of subsets of the data that show an interesting (\ie, significantly different) distribution with respect to a predefined target concept from a large set of candidates.
In our scenario, we perform a series of subgroup searches, each using one survey answer option as the target.
To create the search space of candidate subgroup descriptions, we use all features described in~\Secref{sec:features}.
For the topic features, we consider a topic as present in an article viewed by a user if our topic model provided a probability for this topic above $20\%$.
Other numeric features are binarized in five intervals
using equal-frequency discretization. Due to missing values and multiple occurrences of values, bin sizes can significantly deviate from 20\% of the dataset for some features.
To select the most interesting subgroups, we use the \emph{lift} as a quality function \cite{geng2006interestingness}. 
This measure is computed as the ratio between the likelihood of a survey answer in the subgroup and the respective likelihood in the overall dataset.
As an example, a lift of $1.3$ means that the respective survey answer is $30\%$ more likely to occur in the subgroup than in the overall data.
Additionally, we apply a filter to remove all subgroups that could not be shown to be significant by a $\chi^2$ test with a Bonferroni\hyp corrected threshold of $\alpha = 0.05$.

As a result, we obtain a list with the top $k$ interesting subgroups for each survey answer $T$.
For each subgroup $S$, we can compute various statistics: 
the (relative) \emph{size} $P(S)$ of the subgroup, \ie, the share of users that are covered by the subgroup description, 
the share $P(S|T)$ of subgroup users among those who answered with $T$ in the survey,
the \emph{target share} $P(T|S)$ in the subgroup, \ie, the share of users within the subgroup that reported the respective answer, and the \emph{lift,} which is defined as 
$P(T|S)/P(T) = P(S|T) / P(S)$.
Note that the absence of a feature in the discussion does not mean that it was not considered, but that it is not among the most significant subgroups.

\para{Motivation. }
We start with characterizing groups with specific \emph{motivations} as reported in the survey.
In particular, we provide detailed results for two exemplary motivational groups (work/school and bored/random; \Tabref{tab:sgResults}) and only shortly summarize results for other motivations.


Users who intend to use Wikipedia for work or school are more frequently observed for specific topics of articles, namely
war \& history, mathematics, technology, biology \& chemistry, and literature \& arts.
For the first two of these topics, users are more than twice as often motivated by work or school tasks as on average.
While these topics cover a wide range of different areas, all of them are more related to academic or professional activities than for leisure.
Additionally, this type of motivation is more often reported by users accessing Wikipedia's desktop version. This could be expected since many work\slash school activities are performed in office settings.
Furthermore, we can see that this motivation occurs more often for users who are referred by external search engines multiple times in a session, and by users who stay longer on an individual page, which can be seen as a potential indicator for intensive studying.

By contrast, users who describe their motivation as\linebreak bored/random, are more likely to use internal navigation within Wikipedia and to spend only little time on the individual articles.
Also, they tend to switch topics between the individual articles more often (as indicated by the subgroup with a high average topic distance).
These are telltales for less focused browsing behavior.
Bored users also view more articles on Wikipedia both within the survey session and overall during the respective week.
Finally, this motivation can also be observed more frequently for articles that cover specific topics, such as sports, 21st century, and TV, movies,~\& novels. Clearly, these topics are more leisure\hyp oriented and are in stark contrast to the previously discussed topics favored by users who use Wikipedia for work or school.

Due to limited space, we only outline findings for other motivations:
For example, motivation via media is significantly more often observed for the topics TV, movies,~\& novels (lift $1.37$) and 21st century (lift $1.26$), for popular articles, \ie, articles with a high number of pageviews (lift $1.17$), and for articles in the periphery of the Wikipedia link network according to pagerank (lift $1.14$).
The motivation of looking up something that came up in a conversation is more frequently reported for users with a single Wikipedia article request within a session (lift $1.08$) and for users of the mobile version of Wikipedia (lift $1.08$).
The current\hyp event motivation is more likely for articles about sports (lift $1.97$), 21st century (lift $1.49$), and education, government,~\& law (lift $1.49$). It is also more common for articles with many page views (lift $1.68$), possibly because articles on current events are trending.
Users who aim at intrinsic learning show a topic preference for more scholarly topics such as literature \& art  (lift $1.30$), mathematics (lift $1.24$), and technology (lift $1.21$).
Finally, the geographical origin of a user also has an effect:
the motivations personal decision, current event, and intrinsic learning are reported significantly more often for users from Asia (mostly India; lifts $1.46$, $1.44$, and $1.20$).

\para{Information need. }
Overall, the investigated subgroups are more homogeneous with respect to the reported information need.
We can, however, find some notable (anecdotal) exceptions:
Users from Asia describe their \emph{information needs} significantly more often as acquiring in-depth information (lift $1.51$).
For users who want to obtain an overview of a topic, using the desktop version of Wikipedia is more common than for the average user (lift $1.13$)
Also, topics play a certain role: fact look-ups, for example, are more often observed for the sports topic (lift $1.08$).
Session features that describe user behavior across multiple page visits do not lead to any significant differences in information need.

\para{Prior knowledge. }
Regarding readers' \emph{prior knowledge,} we can observe that users feel familiar with topics that are more spare-time oriented, such as sports (lift $1.21$), 21st century (lift $1.08$), and TV, movies,~\& novels (lift $1.07$).
They also feel more familiar about articles that are popular, \ie,  have many pageviews (lift $1.11$), are longer (lift $1.10$), and are more central in the link network (out-degree, in-degree, or pagerank; lifts $1.11$, $1.09$, and $1.08$).
Naturally, the answer ``unfamiliar'' is more often reported for the exact opposite of these subgroups.
Features that describe a user behavior over multiple article views do not lead to significant deviations.

\subsection{Summary of results}

\para{Prevalence of use cases.}
We have shown that Wikipedia is read in a wide variety of use cases that differ in their motivation triggers, the depth of information needs, and readers' prior familiarity with the topic. There are no clearly dominating use cases, and readers are familiar with the topic they are interacting with as often as they are not. Wikipedia is used for shallow information needs (fact look-up and overview) more often than for deep information needs. While deep information needs prevail foremost when the reader is driven by intrinsic learning, and fact look-ups are triggered by conversations, we saw that overviews are triggered by bored/random exploration, media coverage, or the need for making a personal decision. 

\para{Use cases over time.}
Motivations appear to be mostly stable over time (days of the week and hours of the day), with a few exceptions: motivations triggered by the media are increased over the weekends and at nights, conversation triggers are increased over the weekends, and work/school triggers are increased on week days and during the day.

\para{Behavioral patterns.} By connecting survey responses with webrequest logs, we identified certain behavioral patterns:

\begin{itemize}[leftmargin=*, topsep=0pt,itemsep=-1ex,partopsep=1ex,parsep=1ex]
    \item When Wikipedia is used for work or school assignments, users tend to use a desktop computer to engage in long pageviews and sessions; sessions tend to be topically coherent and predominantly involve central, ``serious'' articles, rather than entertainment\hyp related ones; search engine usage is increased; and sessions tend to traverse from the core to the periphery of the article network.   
    \item Media-driven usage is directed toward popular, entertainment\hyp related articles that are frequently less well embedded into the article network.
    \item  Intrinsic learning tends to involve arts and science articles with no significant navigational features; conversations bring infrequent users to Wikipedia, who engage in short interactions with the site, frequently on mobile devices. 
    \item People who use Wikipedia out of boredom or in order to explore randomly tend to be power users; they navigate Wikipedia on long, fast-paced, topically diverse link chains; and they often visit popular articles on entertainment-related topics, less so on science-related topics. 
    \item Current events tend to drive traffic to long sports and politics-related articles; the articles tend to be popular, likely because the triggering event is trending.
\item When Wikipedia is consulted to make a personal decision, the articles are often geography and technology-related, possibly due to travel or product purchase decisions.
\end{itemize}

\section{Discussion}
\label{sec:discussion}

Every day, Wikipedia articles are viewed more than 500 million times, but so far,
very little has been known about the motivations and behaviors of the people behind these pageviews.
The present study is the first comprehensive attempt to help us understand this group of users by combining a survey with a log\hyp based analysis.

The work most closely related to ours is by Lehmann \etal\ \cite{lehmann2014reader}, who extracted \WP navigation traces from Yahoo!\ toolbar logs (which may be considered a biased sample of the complete logs we have access to) with the goal of discovering a set of usage patterns according to which articles are consumed.
Using clustering techniques, they concluded that there are four types of articles:
trending articles, articles read in a focused manner, articles read by exploring users, and articles users just quickly pass through.
Lehmann \etal's work is entirely ``unsupervised'', in the sense that they have no ground truth of the actual underlying user motivations and needs.

We, on the contrary, have elicited the ground truth through our survey and can thus arrive at stronger and more actionable conclusions, which we discuss next.
We do so by first highlighting implications and directions for future work (\Secref{sec:Implications and future directions}), and then reflecting on our methodology and pointing out its limitations (\Secref{sec:Reflections on methodology}).

\subsection{Implications and future directions}
\label{sec:Implications and future directions}
This research has already had considerable impact within the Wikimedia Foundation, where it has informed several items on the product development agenda, and we hope that it will further inspire Wikimedia developers, academic researchers, and volunteers to build tools for improving the user experience on \WP.


\para{Predicting motivation and desired depth of knowledge.}
A tool immediately suggested by our results could involve statistical models for real-time inference of user session motivations from behavioral traces as captured in the webrequest logs.
Such models could be trained in a supervised fashion with features of \Secref{sec:webrequest_logs} as input, and survey responses as output, and could form the basis for products and services for supporting the needs of \WP readers more proactively.
For instance, if an editor working on an article could be shown an estimate of the distribution of the motivations and desired depths of knowledge on behalf of the readers of the article, she can take this information into account to tailor the content to the needs of the audience or attempt to change the distribution of the audience's motivation by creating specific types of content in the article. 
Such a tool could have large impact, considering that, currently, editors contribute to content on \WP without much knowledge of the users who will eventually read it.

Similarly, predicting the distribution over depths of knowledge sought by the readers of an article could offer opportunities for creating different versions the article, \eg, for those who are interested in quick look-ups \vs\ in-depth readers.
This could enhance the usability of Wikipedia articles particularly on mobile devices with smaller screens and low\hyp bandwidth connections.

The above task of using digital traces to predict survey responses has been called \emph{amplified asking,} and it is known to be difficult \cite{salganik2017bit}.
This has been confirmed by our preliminary attempts, where we have achieved accuracies only slightly better than simple baselines.
This may be partly explained by the fact that user motivations may change during a session, and while the survey captures the motivations at the article level accurately, it fails to capture possible transitions between motivations during a session.
For instance, a session might start with a school or work project in mind, but the user might then transition to procrastinating by exploring \WP randomly, which would not be captured in our current setting.
Also, prediction is complicated by the fact that, even for a fixed article, user motivations might vary widely.
For instance, of the $222$ users taking the survey upon reading the article about Donald Trump, $38\%$ read the article out of boredom, $32\%$ in response to media coverage, $24\%$ because of a conversation, $23\%$ due to current events, $17\%$ because the topic was personally important to them, \etc


Despite these difficulties, future work should investigate the problem of predicting user intentions in more depth.

\subsection{Methodological limitations}
\label{sec:Reflections on methodology}

We discuss certain limitations of present research next.

\xhdr{Survey selection bias}
A general caveat with surveys is that one typically cannot guarantee that whether a subject participates or not is a fully random choice.
Certain covariates may be associated with both participation rates and responses given, leading to biased conclusions.
We made a best effort to correct for this bias by adjusting survey responses based on a random sample of all \WP pageviews drawn from \WP's webrequest logs (\Secref{sec:correction}).
However, if the bias\hyp inducing covariates are hidden, one cannot fully correct for the bias.
For instance, young users might be both more prone to use \WP for work or school and to participate in our survey; this would over\hyp represent the work\slash school motivation in our raw survey results, and since we have no information about users' age, we could not correct for this bias.
Apart from that, survey answers might be biased by social desirability~\cite{demaio1984social};
\eg, even in an anonymous survey, users might be reluctant to admit they are visiting Wikipedia out of boredom.


\xhdr{Unique visitors and level of analysis}
Wikipedia does not require users to log in, nor does it use cookies in webrequest logs to maintain a notion of unique clients.
Hence, we need to rely on an approximate notion of user IDs based on IP addresses and browser versions (\Secref{sec:webrequest_logs}), which makes the attribution of pageviews to users and the construction of sessions imperfect.
In particular, we might not recognize that two pageviews are by the same user if they use several devices or if their IP address changes for other reasons; and we might conflate several users if they share the same device or IP address (\eg, via a proxy).
Currently, we limit the impact of such errors by 
analyzing the data on a session level and
operating at relatively short time scales (an inactivity of more than one hour ends the session being studied).
If the attribution of pageviews to unique users becomes more precise in the future, we could study user behavior at longer time scales, which would, \eg, allow us to understand and support long-term learning needs.
Also, our current method aims at giving each user session equal weight. An alternative approach would be to analyze the data on a request level, which would put more emphasis on the motivations and needs of power users.

\xhdr{Cultural issues}
The results discussed here pertain to the English edition of Wikipedia.
Even within this limited scope, our behavioral analysis hints at subtle cultural and geographical differences; \eg, the bored\slash random motivation is particularly frequent in the U.S., whereas current events are a stronger motivator in India.
Survey answers might also be influenced by different notions and associations of the survey phrasing across cultures \cite{harkness2003cross}.
Since \WP strives to reach beyond cultural and linguistic boundaries, it is important to further investigate these cultural issues.
As part of this effort, we are planning to repeat our study in additional language versions of \WP to elicit cultural differences on a larger scale.

\section{Conclusions}
\label{sec:conclusions}
In this work, we study why users read Wikipedia.
We use survey data to develop a taxonomy of Wikipedia usage along three dimensions: \emph{motivation}, \emph{information need}, and \emph{prior knowledge}.
In a large-scale survey with almost 30,000 participants, we quantify the share of readership for these driving factors.
The bias\hyp corrected survey results reveal a broad range of usage scenarios, interdependencies between survey answers, and temporal trends.
Combining the survey responses with webrequest logs allows us to  characterize motivational groups with behavioral patterns.
The outcomes of this study are currently being discussed in the Wikimedia Foundation as a stimulus for developing specialized tools for readers and editors.
\section*{APPENDIX: SURVEY BIAS CORRECTION}
\label{ap:bias}

This appendix covers the details of the survey bias correction.

\para{Propensity score weight adjustment.}
We use \emph{inverse propensity score weighting} to adjust for potential biases in survey response data with respect to control data \cite{austin2011introduction,lunceford2004stratification}.
Specifically, we want to infer unbiased estimates of survey answers for the whole Wikipedia readership. Thus, we randomly sampled a large set of Wikipedia readers (25 times the number of survey responses) from the webrequests logs in the survey period. Then, we proceeded to sample one request for each selected user and marked it as an imaginary request reflecting a potential survey response; we also deduced the same set of features as for our survey (except responses). We only sampled requests that are desktop or mobile pageviews in English Wikipedia's main namespace and applied bot-filtering in order to match the original survey.

The propensity score of a single instance then reflects the probability that an instance with these control features (\Secref{sec:features}) participated in the survey. We approximate it using our control group.
For that, a post-stratification approach \cite{lunceford2004stratification} is infeasible due to the large number of control features we consider.
Instead, we model the group membership (survey participant or control group) using gradient boosted regression trees showing promising results in the past in comparison to traditional approaches like logistic regression \cite{lee2010improving}.
Given the features of an instance $x$, the model predicts a probability $p(x)$ that $x$ belongs to the survey group.
We then set the weight $w$ for instance to
$1/p(x)$.
The rationale behind this procedure is that answers of users that are overall less likely to participate in the survey receive higher weights since they represent a larger part of the entire population with similar features.

\para{Evaluating weights.}
To evaluate if applied weighting schemes have the intended correcting effect of making the user survey data more representative for the overall Wikipedia, we resort to two scenarios.

First, we check that the resulting weights do not contain drastic outliers dominating subsequent results, which would warrant so-called \emph{trimming} \cite{lee2011weight}. In that regard, we observe that weights are sufficiently homogeneous distributed with a minimum of $1$, a maximum of $190$, 
a mean of $17.6$, and a standard deviation of $26.9$.


Additionally, we evaluate how well we can recover the mean value of features in the overall population from observed survey response features 
and our weighting scheme.
For that purpose, we compute weighted and unweighted averages of the observed values for the survey users and compare them with the mean of a different random sample as a ground truth.
As a result, the average of relative errors is reduced by $86\%$, from $0.556$ in the unweighted case to $0.079$ in the weighted case.
The reduction is strongly significant ($p \leq 0.001$ according to a Wilcoxon signed rank test).
If weighting is applied, then the mean recovered from the weighted survey is never more than $0.2$ standard deviations off compared to the actual feature mean in the sample.





\para{Effective sample size.}
Int this work, we employ a variety of statistical techniques on the survey data. Yet, the introduction of sample weights for correcting bias in survey responses leads to violations of IID assumptions \cite{mukhopadhyay2016complex}.  
Thus, standard errors of estimators are estimated as too small, which in turn leads to confidence intervals being too narrow and statistical tests asserting significance too often if standard procedures are applied. 
The extent to which the sampling error in the survey for some parameter $\theta$ deviates from the expected error from an IID sample due to survey design and correction, is known as the \emph{design effect} (deff) \cite{kish1965survey}. 
If the design effect deviates from $1$---as it is the case in our survey---then our understanding of sample size for calculating standard errors becomes incorrect. 
To that end, we consider the \emph{effective sample size} estimating the required sample size of a random sampling survey for achieving the same error as the weighted sample---it is defined as $n_{\text{eff}}=n/\text{deff}$. As we cannot directly calculate $\text{deff}$ without knowing the expected sampling error, we use Kish's approximation formula with weights $w_i$ \cite{kish1965survey}:

\begin{equation*}
   n_{\text{eff}} =
   \frac{\left(\sum_{i=1}^n w_i\right)^2}{\sum_{i=1}^n w_i^2} 
\end{equation*}

For our complete survey data, $n_{\text{eff}}=8839$.
We use this effective sample size throughout this article for calculating standard errors, confidence intervals, and statistical tests. Note that this makes reported confidence interval and statistical hypothesis tests overly careful. For further details, please refer to \cite{mukhopadhyay2016complex}.

\para{Acknowledgements.}
We thank Dario Taraborelli from Wikimedia Foundation who was indispensable to the early phases of the project. We also thank Jonathan Morgan for helping us with the hand-coding; Jon Katz and Toby Negrin for helping us shape the direction of the research and supporting us throughout; Anne Gomez, Jeff Hobson, Bahodir Mansurov, Jon Robson, and Sam Smith for running the surveys on Wikipedia; and Aeryn Palmer for creating the privacy statements for this research. This research has been supported in part by NSF IIS-1149837, ARO MURI, DARPA NGS2, and Stanford Data Science Initiative.
%
%







%

\clearpage
\begingroup
\small
\raggedright
\sloppy
\balance
\bibliographystyle{abbrv}
\DeclareRobustCommand{\VAN}[3]{#3}
\bibliography{bib}  
\endgroup
%


\end{document}